\documentclass[aps,pre,floatfix,superscriptaddress,showpacs,twocolumn]{revtex4}

\usepackage{graphicx,psfrag}
\usepackage[hypertex]{hyperref}

\newcommand{\average}[1]{\left<{#1}\right>}

\newcommand{\by}{\mathbf{y}}
\newcommand{\bB}{\mathbf{\underline B}}
\newcommand{\bF}{\mathbf{F}}

\newcommand{\En}{\mathcal{N}}

\newcommand{\aut}{\psi}

\newcommand{\p}[1]{\left({#1}\right)}
\newcommand{\pq}[1]{\left[{#1}\right]}

\begin{document}

\title{Work and heat probability distribution
  of an optically driven Brownian particle: Theory and experiments}

\author{A. Imparato}
\affiliation{Dipartimento di Fisica and CNISM, INFN Sezione di Torino, Politecnico di Torino,  C.so Duca degli Abruzzi 24, 10121 Torino, Italy}
\author{L. Peliti}
\affiliation{Dipartimento di Scienze Fisiche, Universit\`a
``Federico II'', Complesso Monte S. Angelo, 80126 Napoli, Italy}
    \affiliation{CNISM, Napoli, Italy}
    \affiliation{INFN, Sezione di Napoli, Italy}
\author{G. Pesce}
    \affiliation{Dipartimento di Scienze Fisiche, Universit\`a
``Federico II'', Complesso Monte S. Angelo, 80126 Napoli, Italy}
    \affiliation{CNR-INFM Coherentia, Napoli, Italy}
\author{G. Rusciano}
    \affiliation{Dipartimento di Scienze Fisiche, Universit\`a
``Federico II'', Complesso Monte S. Angelo, 80126 Napoli, Italy}
    \affiliation{CNISM, Napoli, Italy}
\author{A. Sasso}
    \affiliation{Dipartimento di Scienze Fisiche, Universit\`a
``Federico II'', Complesso Monte S. Angelo, 80126 Napoli, Italy}
    \affiliation{CNISM, Napoli, Italy}

\date{September 15,2007}

\begin{abstract}
We analyze the equations governing the evolution of distributions 
of the work and the heat exchanged
with the environment by a manipulated stochastic system,
by means of a compact and general derivation.
We obtain explicit solutions for these equations for the case
of a dragged Brownian particle in a harmonic potential. We
successfully compare the resulting predictions with
the outcomes of experiments, consisting in dragging a micron-sized
colloidal particle through water with a laser trap.

\end{abstract}
\maketitle

The study of the physics of small systems has recently received a
boost by the possibility of manipulating nanosystems and
biomolecules. The fluctuations of the work and heat that these small
systems exchange with the environment while being manipulated can be
of the order or even larger than the thermal energy, leading to ``transient'' violations of the second principle of  thermodynamics.
The distributions of heat and work have been experimentally studied
for a few brownian systems
\cite{EvansExp02,EvansRev04prl,gar_cil_tot,blickle_06}.
The probability distribution function (PDF) of the work done on a
Brownian particle dragged by a moving quadratic potential was
derived in \cite{maz_jarz_99,zon_cohen_04}. The distribution
turns out to be gaussian, what has been taken
as an ansatz in \cite{zon_cohen_04} and confirmed
in \cite{tani_cohen_07} by
means of a rather involved path integral calculation. On the other
hand, obtaining the PDF of the transferred heat represents 
a much more difficult
task: the Fourier transform of this function was obtained in
refs.~\cite{zon_cohen_04,tani_cohen_07} by exploiting the energy
balance and the gaussian ansatz for the work PDF, valid when the
potential is quadratic.

In the present paper we derive in a simple way the differential
equations governing the evolution of the PDFs of the work and heat
exchanged by a brownian particle, valid for any choice of the
potential acting on the particle. The solutions of these equation
turn out to fulfill the well-known fluctuation relations. We
evaluate the solution of these equations for a moving harmonic
potential. We then experimentally study the work and the heat
exchanged by a colloidal particle dragged through water by an
optical trap. The PDF's predicted by our equations
result in an excellent agreement with the experimental
data.
We were inspired by the experiment of Wang {\it
et~al.}~\cite{EvansExp02} where the work done on a similar system
was measured. However, in that experiment, only the performed work,
and not the heat transferred, was sampled. Moreover, the expected
gaussian distribution of the performed work was not verified, and
a detailed comparison with the theoretical predictions was not
attempted. However in a subsequent paper \cite{reid04},
 the authors stressed that the  PDF of the work has to be gaussian in their experimental conditions.

Let us consider a Brownian particle in the overdamped regime, driven
by a time-dependent potential $U(x,X(t))$, where $X$ is an
externally controlled parameter, that varies according to a fixed
protocol $X(t)$. The Langevin equation is given by
\begin{equation}
\frac {dx} {dt}=-\Gamma U'(x,X)+f(t),
\label{eqLa}
\end{equation}
where $\average{f(t) f(t')}=(2\Gamma/\beta) \delta(t-t')$ and
the prime denotes derivative with respect
to $x$.  We have defined $\beta=(k_\mathrm{B}T)^{-1}$, and
$\Gamma=1/6\pi r \eta$, for a spherical particle with radius
$r$, in a medium of viscosity  $\eta$.

The {\it thermodynamical} work done on the particle is defined by
\begin{equation}
W=\int d X\, \frac{\partial U}{\partial X}
=\int_0^{t} d t'\, \dot X(t') \frac{\partial U}{\partial X} .
\label{wdef}
\end{equation}
Besides work, the particle also exchanges with the environment heat,
whose expression is
\begin{equation}
Q=\int d x\, \frac{\partial U}{\partial x}=\int_0^t d t'\, \dot x(t') U'(x(t'),X(t')).
\label{qdef}
\end{equation}
If $Q>0$ the particle receives heat by the environment. Note that
the integrals appearing in the last equation are stochastic
integrals that must be interpreted according to the Stratonovich
integration scheme \cite{Sekimoto97}. Let $\Delta U$ be the potential energy
difference between the final and the initial state: the balance
of energy for the manipulated
particle reads $\Delta U = W+ Q$, which follows immediately from
eqs.~(\ref{wdef},\ref{qdef}). Note that the quantities $Q$ and $W$ have to be regarded as stochastic variables, whose value at time $t$ depends on the specific stochastic trajectory.

The differential equation governing the time evolution of the
PDF of the work is given by
\cite{Imparato05a+b}
\begin{equation}
\partial_t\phi(x,W,t)=\Gamma \frac{\partial}{\partial x}
\left[ U'\,\phi\right]+\frac{\Gamma}{\beta}
\frac{\partial^2 \phi}{\partial x^2}
-\dot X\frac{\partial U}{\partial X}  \frac{\partial \phi}{\partial W}.
\label{phi_eq}
\end{equation}
It can be easily shown that the solution of eq.~(\ref{phi_eq})
satisfies the Jarzynski equality \cite{Imparato05a+b}.

The differential equation for the joint PDF $\varphi(x,Q,t)$ of the
position $x$ and the heat $Q$ is obtained as follows. In a short
time interval $\delta t$ the heat exchanged by the particle with the
environment reads $\delta Q=d U-\partial_t U\,\delta t=U' d x$, and
thus the time derivative of $Q$ is given by
\begin{equation}
\frac {d Q}{d t}=U'\frac {d x}{d t}=U'\,(-\Gamma U'+f(t)).
\label{eq_qt}
\end{equation}
Thus equations  (\ref{eqLa}) and (\ref{eq_qt}) describe two coupled
stochastic processes. We now define the vectors of the stochastic
variables $\by$ and of the forces $\bF$ by
\begin{equation}
\by=\left( \begin{array}{c} x \\Q \end{array} \right), \quad
\bF=\left( \begin{array}{c} -\Gamma  U' \\-\Gamma U'^2 \end{array}
\right),
\end{equation}
and the diffusion matrix
\begin{equation}
 \bB =\left( \begin{array}{c c} \Gamma/\beta,&  (\Gamma/\beta) U' \\
 (\Gamma/\beta) U' &     (\Gamma/\beta) U'^2 \end{array} \right).
\end{equation}
Then the differential equation governing $\varphi(x,Q,t)$
straightforwardly follows  \cite{Zwa,spe_seif_05b}:
\begin{equation}
\partial_t \varphi(x,Q,t)=
-\frac{\partial}{\partial \by}
\left(\bF \varphi\right)+\frac{\partial}{\partial \by}
\left(\bB \cdot \frac{\partial}{\partial \by}\varphi\right).
\label{fp:eq}
\end{equation}
By introducing the generating function $\chi(x,\lambda,t)=\int d
Q \,\exp(\lambda Q) \varphi(x,Q,t)$, we obtain the simpler
equation
\begin{eqnarray}
\label{gen_q}
\partial_t \chi(x,\lambda,t)&=&\frac{\Gamma}{\beta}
\frac{\partial^2\chi}{\partial x^2}+\Gamma
\left(1-\frac{\lambda}{\beta}\right)\partial_x \p{ U' \chi}\\
&&\quad{-}\lambda \frac{\Gamma}{\beta} U' \partial_x \chi
+  \lambda\left(\frac{\lambda}{\beta}-1\right) \Gamma U'^2\chi, \nonumber
\end{eqnarray}
that was first derived by  Lebowitz e Spohn \cite{LebSpohn99}. Note
that the operator appearing on the rhs of this equation changes into
its adjoint by the substitution $\lambda\longrightarrow
\beta-\lambda$. 
As discussed in ref.~\cite{LebSpohn99}, this symmetry implies 
the Gallavotti-Cohen
fluctuation relation~\cite{GallCoh95} for our system.

By defining the function $g(x,\lambda,t)$ as
$\chi(x,\lambda,t)=g(x,\lambda,t) \exp\left[-\delta(\lambda)
U(x,t)/2\right]$, with $\delta(\lambda)=\beta -2 \lambda$,
eq.~(\ref{gen_q}) becomes
\begin{equation}
\partial_t g=\frac{\Gamma}{\beta} \frac{\partial^2 g}{\partial x^2}
-\Gamma \beta  \frac{U'^2}{4} g +\frac \Gamma 2 U''g
+\frac{\delta(\lambda)}{ 2 }g  \partial_t U. \label{eq_g}
\end{equation}
This equation has the form of an imaginary-time Schr\"odinger
equation.

It is worth remarking that eqs.~(\ref{phi_eq}), (\ref{fp:eq}) and
(\ref{gen_q}) hold for any choice of the potential $U(x,X(t))$.
Moreover, the present approach can be easily generalized to the case
of a Brownian particle with
inertia~\cite{Imparato06,tani_cohen_07,tani_cohen07b}.

We now consider the particular case of a harmonic potential
\begin{equation}
U(x,t)=\frac k 2(x-X(t))^2,
\label{U2}
\end{equation}
with $X(t)=v t$, i.e., the center of the potential moves with a constant velocity $v$. We shall assume that the particle is initially in
thermal equilibrium, with the potential centered at $X=0$ at $t=0$.
It is then possible to solve directly equation~(\ref{phi_eq}),
obtaining
\begin{equation}\label{pxw}
    \phi(x,W,t)=\mathcal{N}_t \exp\left[
    -\frac{\left(W-\widehat W(x,t)\right)^2}{2\sigma^2(t)}
    -\frac{\beta k}{2}\left(x-\xi(t)\right)^2\right].
\end{equation}
In this equation, having defined $\tau=1/\Gamma k$ and $\alpha(t)=e^{-t/\tau}$,
we have $\xi(t)= v\tau\left(\alpha(t)-1+t/\tau\right)$,  $\sigma^2(t)=v^2 \tau^2 k\beta^{-1} [ 2t/\tau+1-\left(2-\alpha(t)\right)^2]$,
\begin{eqnarray}
\widehat W(x,t)&=& t v^2 \tau k (2-\alpha(t))- vx \tau k (1-\alpha(t))\\
&&{}-v^2 \tau^2 k \left(2+\alpha^2(t)-3\alpha(t)\right);\nonumber\\
\mathcal{N}_t^{-1}&=&\sqrt{4(\pi v \tau/\beta)^{2}
 \left(2t/\tau+1-\left(2-\alpha(t)\right)^2\right)}.
\end{eqnarray}
The unconstrained PDF $\Phi(W,t)\equiv\int d x\, \phi(x,W,t)$, is given by
\begin{equation}
\Phi(W,t)=\En_t'\exp\pq{-\beta\frac{   \left(W- v^2\tau^2 k \left(\alpha(t)-1+t/\tau
\right)\right)^2}{4v^2 \tau^2 k\left(\alpha(t)-1+t /\tau\right)}},
\label{pw}
\end{equation}
where $\En'_t=\left[ 4 \pi \beta^{-1} v^2 \tau^2
k\left(\alpha(t)-1+t/\tau \right)\right]^{-1/2}$. A similar result
was first obtained in~\cite{maz_jarz_99} in a special case,
and then in \cite{zon_cohen_04}, by using qualitative
arguments and by {\it assuming} that $\Phi(W,t)$ is gaussian,
and more recently in \cite{tani_cohen_07}, by a functional integral
technique. We now see that
eqs.~(\ref{pxw},\ref{pw}) can be straightforwardly derived as
solutions of eq.~(\ref{phi_eq}). An approach analogous to ours was
used in ref.~\cite{spe_seif_05},  leading again to~eq.~(\ref{pw}).

We now turn to the heat PDF: substituting eq.~(\ref{U2}) into 
eq.~(\ref{eq_g}) one obtains a Schr\"odinger-like
equation for the harmonic oscillator, that can be solved exactly.
Assuming that the particle is at thermal equilibrium at $t=0$, with
$v=0$ for $t<0$, the solution of eq.~(\ref{gen_q}) reads
\begin{eqnarray}
\chi(x,\lambda,t)&=&\exp\pq{-\frac{\delta(\lambda)}{2} U(x,X(t))
-\frac{ \beta v}{2 \Gamma} z(x,t)}\nonumber \\
&&{}\times\sum_{n=0}^\infty e^{\gamma_n t}  c_n(\lambda) \aut_n(z(x,t)),
\label{chi_tot_t_1}
\end{eqnarray}
where $\gamma_n =\left(- n/\tau +\delta^2(\lambda) v^2/4 \Gamma\beta
-\beta v^2 /4 \Gamma \right)$, and $z(x,t)=x-vt +\delta(\lambda)
v\tau/\beta$, and where $\aut_n(z)$ are the eigenfunctions of the
Schr\"odinger equation for the  harmonic oscillator, with the
substitutions $\hbar^2/m\longrightarrow 2 \Gamma/\beta$ and $m
\omega^2 \longrightarrow\Gamma \beta k^2/2$. The value of
coefficients $c_n(\lambda)$ is determined by the initial condition
$\chi(x,\lambda,t=0)$. Note that $\tau=1/\Gamma k$ sets up the
characteristic time scale for both work and heat fluctuations.

\textit{Case a):} $v=0$ for $t\ge0$, i.e., a fixed
potential. The behavior of the generating function
$\Psi(\lambda,t)\equiv\int d x\, \chi(x,\lambda,t)$ in the long-time
limit is governed by the eigenfunction $\psi_0(z(x,t))$ associated
with the smallest eigenvalue. Thus, after some algebra, one finds
$\Psi(\lambda,t\rightarrow \infty)=1/\sqrt{1-(\lambda/\beta)^2}$.
Therefore, in the case of constant potential, the heat unconstrained
PDF in the long time limit has the expression
\begin{equation}
\varphi(Q,t{\rightarrow} \infty)= \int \frac {d \lambda} {2 \pi i}
\Psi(  \lambda,t{\rightarrow} \infty) e^{- \lambda Q}=\beta \frac{K_0(\beta |Q|)} \pi,
\label{eq_bess}
\end{equation}
where $K_0(x)$ is the zero-th order modified Bessel function of the second kind.

\textit{Case b):} $v>0$. Also in this case the long time
behavior of the solution of eq.~(\ref{gen_q}) will be dominated by
the eigenfunction with $n=0$, which is a gaussian function. Thus, in
the long time limit, one finds
\begin{eqnarray}
&&\Psi(\lambda,t\rightarrow \infty)=
\exp\left\{\frac{v^2\beta}{4\Gamma}
\left[\frac{4\lambda}{\beta}\left(\frac{\lambda}{\beta}-1\right) t \right.\right.\\
&&\qquad\left.\left.{}
+\frac{ 2 \lambda (3 -4 (\lambda/\beta)^2)}{\Gamma k (\beta+\lambda)}
\right]\right\} \left[1-\left(\frac{\lambda}{\beta}\right)^2\right]^{-1/2},\nonumber
\end{eqnarray}
and, integrating by the saddle-point method, one finds
\begin{eqnarray}
&&\varphi(Q,t\rightarrow \infty)=\frac 1 {2 \pi \mathrm{i}} \int d
\lambda\,  \Psi(\lambda,t\rightarrow \infty)\, e^{ \lambda
Q}\label{int_Q_t}
\\
&&\quad {}=\exp\pq{-\frac {\Gamma\beta} {4 v^2 t}\p{Q +\frac {v^2 t } \Gamma}^2}
\sqrt{\frac{\Gamma\beta}{ \pi 4 v^2 t}},
\label{Q_t}
\end{eqnarray}
Note that in the long time limit $\overline W+\overline Q=0$. As
discussed in ref.~\cite{zon_cohen_04}, some care has to be taken
when calculating the integral (\ref{int_Q_t}) with the saddle point
method. The resulting calculation shows that the function $\varphi(Q,t)$
is gaussian up to a
subleading term of order $1/\sqrt t$~\cite{zon_cohen_04}.

In order to test these results, we have experimentally observed the
trajectories of a colloidal particle in an optical trap, which is
well described by a quadratic potential (\ref{U2}) near its focus
$X(t)$.
\begin{figure}[ht]
\center
\psfrag{W}[lt][lt][1.]{$W$  (units of $k_B T$)}
\includegraphics[width=8cm]{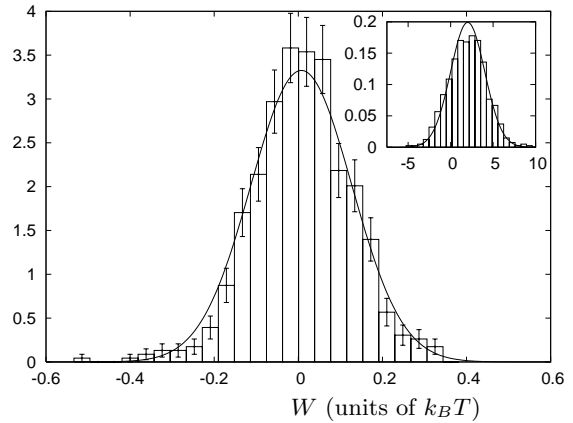}
\caption{Histogram of the work exerted on the colloidal particle by
the optical trap, as given by eq.~(\ref{wdef}), for $t=0.01\, \mathrm s<\tau$ (main
figure) and $t=0.5 \, \mathrm s \gg \tau$ (inset). The lines correspond to the expected
function (\ref{pw}), with no adjustable parameter.} \label{histoW}
\end{figure}
\begin{figure}
\center
\psfrag{Q}[lt][lt][1.]{$Q$  (units of $k_B T$)}
\includegraphics[width=8cm]{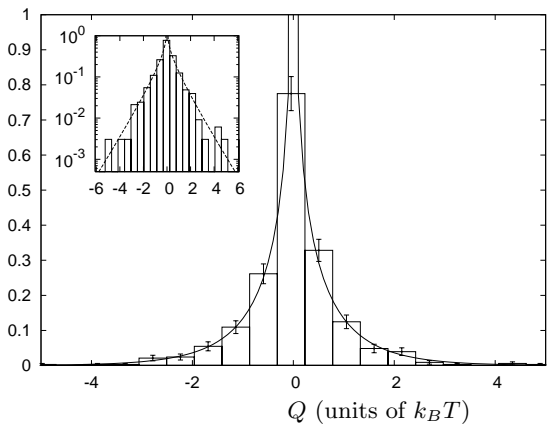}
\caption{Histogram of the heat exchanged by the colloidal particle
with the environment, with fixed optical trap $v=0$, as given by
eq.~(\ref{qdef}), for $t=0.5$ s. The line correspond to the expected PDF
as given by eq.~(\ref{eq_bess}), with no adjustable parameter.
Inset: same data, with logarithmic $y$-axis.} \label{histoQ0}
\end{figure}

The Optical Tweezers system consisted of a home made optical
microscope with a  high numerical aperture water immersion objective
lens (Olympus, UPLAPO60XW3, NA=1.2) and a frequency and amplitude
stabilized Nd-YAG laser ($\lambda=1.064~\mu m$, 500 mW, Innolight
Mephisto). The sample cell was made with a glass coverslip of 150
$\mu$m thickness and a microscope slide glued together by  parafilm
stripes of about 100 $\mu$m thickness. Polystyrene micro-spheres
produced by  Postnova (density: 1.06 g/cm$^3$, refractive index:
1.65) with a diameter of 2.00$\pm0.05~\mu$m  were diluted in
distilled water to a final concentration of about 1$\div$2
particles/$\mu $l. The sample cell was mounted on a closed-loop piezoelectric transducer 
stage (Physik Instrumente PI-517.3CL)  which allowed movements with
nanometer resolution. Moving the stage in a given direction
corresponds to moving the optical trap focus in the opposite
direction. The sample temperature was not stabilized but
continuously monitored using a negative temperature coefficient 
thermistor positioned on the top
surface of the microscope slide. The temperature during a complete
set of measurements remains constant $T=296.5$~K, within 0.2~K. The
trapped bead was positioned in the middle of the sample cell to
avoid any surface effects. The thermally driven motion of trapped
beads was monitored by a  InGaAs quadrant photodiode (Hamamatsu
G6849) placed in the back focal plane of the condenser lens
\cite{GittesOL98}. The response of our quadrant photodiode was linear for
displacements of about 300 nm with a resolution of 2 nm, and its
bandwidth was about 250 kHz. The trajectories in the transverse x-y
plane were sampled at 125 kHz using a digital oscilloscope (details
on the experimental setup can be found elsewhere \cite{PesceRSI05}).
The duration of each trajectory measurement was 10 s. During the
first 5 seconds the stage was at rest and we use this period to
compute the power spectral density of the particle position to
obtain the trap stiffness, which takes the value $k=6.67\times
10^{-7}$ N/m, and the calibration factor of the quadrant
photodiode~\cite{BuoscioloOC04}. Then at $t=5$ s the stage started
to move with a speed of  $v=1 \mu m$/s for 5 s along one axis. After
a pause of 1 second the above described sequence started again, but
the stage was moved in the opposite direction.  We repeated this
procedure 300 times and back and forth trajectories were recorded
for further analysis. Thus the overall number of trajectories
considered is 600. Note that for each trajectory we measure both the
work done on the particle and the heat, as defined by
eq.~(\ref{wdef}) and (\ref{qdef}), respectively.
We have, under our
experimental conditions, $\Gamma=1/6\pi r \eta=5.76 \times 10^7\, \mathrm{s^2/kg}$, yielding $\tau\simeq0.026$ s.

In figure \ref{histoW}, we plot the histogram of the work exerted on
the particle by the optical trap, finding a good agreement with the
expected PDF $\Phi(W,t)$, as given by eq.~(\ref{pw}): the
distribution of the work turns out to be gaussian both at short and
long times. 
At short
times ($t=0.01$~s) the gaussian is peaked around zero, but the peak
moves to positive values of $W$ as $t$ increases.
\begin{figure}
\center
\psfrag{Q}[lt][lt][1.]{$Q$  (units of $k_B T$)}
\includegraphics[width=8cm]{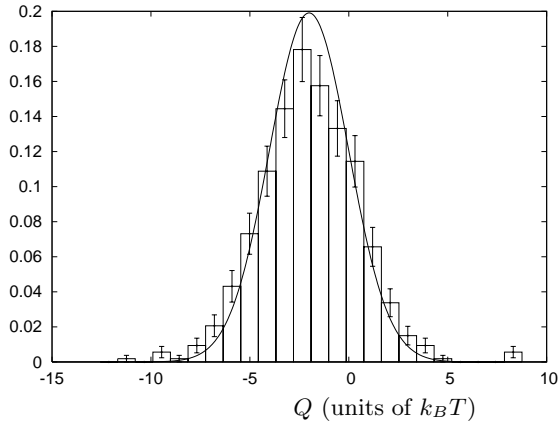}
\caption{Histogram of the heat exchanged by the colloidal particle
with the environment, with $v=1\, \mu \mathrm {m/s}$, and $t=0.5 \, \mathrm s \gg \tau$, as given by
eq.~(\ref{qdef}). The line correspond to the expected PDF as given
by eq.~(\ref{Q_t}), with no adjustable parameter.} \label{histoQ}
\end{figure}

In order to evaluate the heat $Q$ from the particle trajectories, we
exploit the discrete version of eq.~(\ref{eq_qt}): $Q=\sum_{t_i} \left(x_{t_i}
-x_{t_{i-1}}\right) k/2 \left[\left(x_{t_i}-X(t_i))+(x_{t_{i-1}}-X(t_{i-1})\right)\right]$. We
plot the histogram of the measured heat for the motionless trap, with $t=0.5$ s,
 in fig.~\ref{histoQ0}. The histogram agrees nicely with
eq.~(\ref{eq_bess}), in particular the tails of the distribution are
found to be exponential. This behavior was also found at shorter
times (data not shown).

Finally, in fig.~(\ref{histoQ}) the histogram of the measured heat,
for the trap moving with $v=1\, \mu\mathrm{m}/$~s, is plotted, in
the long-time range $t=0.5\;\mathrm{s} \gg \tau$. The distribution is
found to be gaussian, in agreement with eq.~(\ref{Q_t}). By
comparing figures~\ref{histoW} and \ref{histoQ}, it can also be seen
that the mean values of the work and of the heat
are the negative of each other,
as expected in the long-time range.
At shorter times we observe that the tails of the distribution
of the measured heat fall off exponentially,
with time-dependent slopes (data not shown).
We have noticed that for observation times longer than 0.5 s both the heat
and work distributions appear slightly broader than the theoretical predictions,whereas  their centers remain in good agreement with the expected ones (data not shown).
We ascribe this fact to the presence of low-frequency  
(smaller than 1 Hz) noise  affecting our experimental set-up.

We have shown that it is possible to solve explicitly the
differential equations governing the evolution of the PDF for the
work and heat exchanged by a dragged brownian particle, and that the
resulting predictions are vindicated by experiment. In particular
one observes a non negligible probability for the ``transient
violations'' of the second law of thermodynamics, i.e., positive
values of  the exchanged heat $Q$. 

\bibliography{fluctheor,peppe}
\end{document}